\documentstyle[12pt,aaspp4]{article}  
\begin{document}

\def\ra{$\rightarrow$}

\title{Small Deviations from Gaussianity and\\
The Galaxy Cluster Abundance Evolution}

\author{A. L. B. Ribeiro\altaffilmark{1}, 
C.A. Wuensche\altaffilmark{2} and
P. S. Letelier\altaffilmark{1}
\altaffiltext{1}{Departamento de Matem\'atica Aplicada - IMECC,
Universidade Estadual de Campinas, 13083-970 SP, Brazil}
\altaffiltext{2}{Divis\~ao de Astrof\'{\i}sica - INPE, 12201-970 SP,
Brazil}}

\begin{abstract}

In this work, we raise the hypothesis that the
density fluctuations field which originates the growth of
large scale structures is a combination of two or more distributions, instead of assuming the  widely accepted idea that the observed distribution of matter stems from a single Gaussian field produced in the very early universe. By applying the statistical analysis of 
finite mixture distributions to a specific combination of 
Gaussian plus non-Gaussian random fields, we studied the case
where just a small departure from Gaussianity is allowed.
Our results suggest that even a very small level of
non-Gaussianity may introduce significant changes in
the cluster abundance evolution rate.

\end{abstract}

\keywords{Cosmology: large-scale structure of the universe -- cosmology: theory --- galaxies: clusters}
 
\section{Introduction} 

Generally, the problem of structure formation is
associated to the gravitational growth of small density
fluctuations generated by physical processes in the very early universe.
Also, these fluctuations are supposed to build a Gaussian 
random field (GRF), where the Fourier components $\delta_k$
have independent, random and uniformly distributed phases.
Such a condition means that phases are non-correlated in space and
assures the statistical properties of the GRF
are completely specified by the two-point correlation function or,
equivalently, by the power spectrum
$P(k) = |\delta_k|^2$, which contains
information on the density fluctuation amplitude of each scale $k$. 
This  makes the choice of a GRF the simplest initial condition 
for structure formation studies from the
mathematical point of view. At the same time,
the GRF simplicity is  vindicated by a great number of
inflationary models that predict a nearly scale-invariant spectrum
of Gaussian density perturbations from quantum-mechanical fluctuations
in the field that drives inflation (Guth \& Pi 1982).
Likewise, the central limit theorem guarantees a GRF if a wide range 
of random physical processes acts on the distribution of matter in the early universe.

However, a number of mechanisms can generate non-Gaussian
density fluctuations. For instance, they arise
in some inflation models with multiple scalar fields (e.g. Salopek,
Bond \& Bardeen 1989); or after phase transitions when different types of topological defects 
can be formed (Kibble 1976); still, by any discrete, random distributed seed
masses like primordial black holes and soliton objects
(Sherrer \& Bertschinger 1991); as well as in astrophysical processes
during the non-linear regime where early generations of massive stars produce shocks which  sweep material on to giant 
blast waves triggering formation of large-scale structure (Ostriker \& Cowie 1981). Thus, in order to better understand the process of structure formation, it is necessary to investigate the possibility of the
non-Gaussian statistics contribution to the density fluctuation field.

Due to the difficulty to work with generic statistical models, the 
usual approach is to examine specific classes of non-Gaussian 
distributions. Examples of these efforts are the studies
carried out by Weinberg \& Cole (1992)
that studied non-Gaussian initial conditions generated by a range
of specific local transformations of an underlying Gaussian field;
Moscardini et al. (1991) investigated whether non-Gaussian
initial conditions can help to reconcile the CDM models with observations; and Kayama, Soda \& Taruya (1999), who used data on the abundance of clusters at three different redshifts to establish constraints on structure formation models based on  chi-squared non-Gaussian fluctuations generated during inflation. 

In this work, we propose a new approach to this problem,
exploring the hypothesis that initial
conditions for structure formation do not build a single GRF, but a combination of different fields, produced by different
physical mechanisms, whose  resultant effect presents an arbitrarily small  departure from the strict Gaussianity.  The paper is organized as follows:
in Section 2, we introduce the statistical
analysis of finite mixture distributions and
present a two-component mixture model;
in Sections 3, we apply the model to the cluster abundance
evolution; in Section 4 we summarize and discuss our results.

\section{Mixture Distributions Models: The Positive Skewness Case}

Suppose the density fluctuations field, given by
the density contrast $\delta = (\rho(r) - \overline{\rho})/\overline{\rho}$,
is a random variable which takes values in a sample space $\Re$,
and that its distribution can be represented by a probability density
function of the form
$$p(\delta)~=~\alpha_1f_1(\delta)+...+\alpha_kf_k(\delta)~~~~~(\delta\in\Re)\eqno(2.1)$$
\noindent where
$$\alpha_j>0, ~~~~~j=1,...,k; ~~~~~~ \alpha_1+...+\alpha_k=1$$
\noindent and
$$ f_j(\delta) \geq 0, ~~~~~~ \int_{\Re}f_j(\delta)\,d\delta~=~1,~~~~
j=1,...,k.$$

\noindent When this happens, we say that $\delta$ has a finite mixture distribution
defined by (2.1), where 
the components of the mixture are $f_1(\delta),...,f_k(\delta)$ and
the mixing weights are $\alpha_1,...,\alpha_k$
(e.g. Titterrington, Smith \& Makov 1985). Note that we are not using
here the central limit theorem. Mathematically, this will be  valid only when $k\rightarrow\infty$ and the weights have similar values, so that
one process has no more importance than the others. We are not making
these hypotheses here and, consequently, the summation of processes will
not necessarily converge to a Gaussian.

Statistical evidence for a small level of non-Gaussianity in the anisotropy of the cosmic background radiation temperature has been found in the COBE 4 year maps (e.g. Ferreira, Magueijo \& G\'osrki 1998; Pando,
Valls-Gabaud \& Fang 1998; Magueijo 1999). 
Non-Gaussian statistics is also expected in the framework of biased models of galaxy formation (Bardeen et al. 1986). In this case, analytical arguments show that non-Gaussian behaviour corresponds
to a threshold effect superimposed on the Gaussian background
(Politzer \& Wise 1984; Jensen \& Szalay 1986).
In the same way, hybrid models
show that it is possible for structure to be seeded by a weighted combination of adiabatic perturbations produced during inflation and  active
isocurvature pertubations produced by topological defects generated at
the end of the inflationary epoch (e.g. Battye \& Weller 1998).
Thus, a very compelling way to simplify our model is to apply (2.1) to the combination of only two fields: a GRF plus a second field, where the latter will represent a small departure from the strict
Gaussianity. This can be posed as
$$p(\delta)~=~\alpha f_1(\delta)+(1-\alpha)f_2(\delta)\eqno(2.2)$$

\noindent The first field will be always the Gaussian component and
a possible effect of the second component is to modify the 
GRF to have  positive and/or negative tails.
The parameter $\alpha$ in (2.2) allow us to modulate the relative importance
between the two components. It represents an arbitrarily small departure from the strict Gaussianity and can be due to some primordial mechanism acting on the energy distribution.  Such a two-component random field can be generated by taking $\delta_k^2=P(k)\nu^2$, where $\nu$ is a random number with distribution given by (2.2). Then we have
$$\langle \delta^2(r) \rangle = {V\over (2\pi)^3} \int_k P(k)
\left[\int_\nu [\alpha f_1(\nu) + (1-\alpha)f_2(\nu)]\nu^2d\nu\right]d^3k\eqno(2.3)$$

\noindent where $V$ is the volume of an arbitrarily large region of the universe. The 
quantity in the brackets will be defined as the mixture term
$$T^{mix}\equiv \int_\nu [\alpha f_1(\nu) + (1-\alpha)f_2(\nu)]\nu^2d\nu\eqno(2.4)$$

\noindent so that $P(k)^{mix}\equiv P(k)T^{mix}$, for the case where $\alpha$ is not scale-dependent. In the same way, the $rms$ mass overdensity within a certain scale $R$ will be $\sigma^2(R)^{mix}\equiv \sigma^2(R) T^{mix}$.

As an ilustration, in this work we explore 
the case of a positive skewness model, 
where the second field adds to the Gaussian component a positive tail representing a number of rare peaks in the density fluctuation field. 
A simple way to obtain this effect is to take the well known
lognormal distribution as the second component. 
Besides its mathematical simplicity, this distribution seems
to play an important role over the non-linear regime for a wide
range of physical scales (e.g. Coles \& Jones 1991, Plionis \& Vardarini 1995, Bi \& Davidsen 1997). Accordingly, our mixture becomes
$$f_1(\nu) = {1\over{\sqrt{2\pi}}}e^{-{1\over2}\nu^2}~~~{\rm and}~~~
  f_2(\nu) = {1\over{\nu\sqrt{2\pi}}}e^{-{1\over2}(ln\nu)^2}\eqno(2.5)$$

\noindent (for the case of mean zero).
Introducing (2.5) in (2.4), we find
$$T^{mix}= \int_\nu \left[{\alpha\over{\sqrt{2\pi}}}e^{-{1\over2}\nu^2}  + {(1-\alpha)\over{\nu\sqrt{2\pi}}}e^{-{1\over2}(ln\nu)^2}\right]\nu^2d\nu\eqno(2.6)$$

\noindent Resolving this integral we have
$$T^{mix}=\left[\alpha + {e^2\over 2}(1-\alpha)\right]\eqno(2.7)$$

\noindent Hence, if $\alpha\approx 1$, then $P(k)^{mix}\approx P(k)$ and $\sigma^2(R)^{mix}\approx\sigma^2(R)$, which means that a sufficiently small
contribution of the second field leaves the amplitude and shape of the
power spectrum and the mass fluctuation practically unchanged.

\section{Cluster Abundance Evolution}

The correct framework to describe the evolution of non-linear
objects in the context of this model requires a generalization of the
Press \& Schechter formalism (Press \& Schechter 1974) in order to 
take into account the second field.  Assuming that only regions with $\nu > \nu_c$ will form gravitationally bound objects with mass larger than $M$
by the time $t$, the fraction of these objects can be calculated through 
$$F(M)=\int_{\nu_c}^\infty p(\nu)d\nu, \eqno(3.1)$$

\noindent where $\nu=\delta/{\sigma_R}$. 
This quantity is transformed into the comoving number
density of objects with mass
between $M$ and $M+dM$ by taking $\partial F/\partial M$ and dividing
it by $(M/\rho_b)$. Thus, 
$$n(M)dM=2{\left(\rho_b\over M\right)}{\partial\over \partial M}
\left[\int_{\nu_c}^\infty p(\nu)d\nu\right]dM \eqno(3.2)$$

\noindent where $\rho_b$ is the background density and
the number 2 comes from the correction factor $[\int_0^\infty p(\nu)d\nu]^{-1}=2$, which takes into account all the mass of the universe.
If $p(\nu)$ is given by (2.2), then (3.2) can be written as
$$n(M)dM=2{\left(\rho_b\over M\right)}{\partial\over \partial M}
\left[\int_{\nu_c}^\infty [\alpha f_1(\nu) + (1-\alpha)f_2(\nu)] d\nu\right]
dM \eqno(3.3)$$
\noindent Now, introducing (2.5) in (3.3) we have
$$n(M)dM = \sqrt{{2\over \pi}}\left({\rho_b\over M}\right)\left[
\alpha\left({\partial\nu_c\over\partial M}\right) e^{-{\nu_c^2\over 2}}+ 
(1-\alpha)\left({\partial\ln\nu_c\over\partial M}\right) e^{-{(\ln\nu_c)^2\over 2}}\right]dM \eqno(3.4)$$

\noindent Following Sasaki (1994), we rewrite (3.4) to give the density of objects with mass in the range $dM$ about $M$ which virialize 
at the redshift $z$ and survive until the present epoch without
merging with other systems. It becomes
$$n(M,z)=F(\Omega)\left({M\over M_\ast(z)}\right)^{{(n+3)}\over 3}\sqrt{{2\over \pi}}\left({\rho_b\over M^2}\right){{(n+3)}\over 6}
\left[\alpha A(M,z) +(1-\alpha) B(M,z)\right]\eqno(3.5)$$

\noindent where 
$$F(\Omega) = {5\over 2}\Omega\left[{(1+{3\over 2}\Omega)\over
(1+{3\over 2}\Omega +{5\over 2}\Omega z)^2}\right],~~
A(M,z)=\left({M\over M_\ast(z)}\right)^{{(n+3)}\over 6}exp\left[-{1\over 2}\left({M\over M_\ast(z)}\right)^{{(n+3)}\over 3}\right],$$

$$B(M,z)=exp\left[-{1\over 2}\ln\left({M\over M_\ast(z)}\right)^{{(n+3)}\over 6}\right]^2~{\rm{and}}~~ M_\ast(z)=M_\ast(1+z)^{-6/(n+3)}$$

Eq.(3.5) allows us to compare the cluster abundance
evolution with observational data. Clusters, as the most
massive collapsed structures, correspond to rare peaks
in the primordial density field and so their abundance is
sensitive to the occurence of non-Gaussianity in the density fluctuation distribution. Also, cluster evolution provides a constraint on the amplitude
of the mass fluctuation at 8 $h^{-1}$ Mpc scale, $\sigma_8$,
and on the cosmological density parameter, $\Omega_m$, through the relation
$\sigma_8\Omega_m^{0.5}\simeq0.5$ (e.g. Henry \& Arnaud 1991; Pen 1998). 
In a recent work, Bahcall (1999) shows that several independent methods
based on clusters data indicate a low mass densitiy in the universe,
$\Omega_m\simeq 0.2$ and, in consequence, $\sigma_8\simeq 1.2$,
breaking the degeneracy between these parameters.

Here, we compare  the behaviour of the cluster abundance evolution
given by (3.5) with data compiled by Bahcall \& Fan (1998). As an example, we plot in Figure 1a some fits to the observational data for two different values of $\Omega_m$ (0.2 and 1.0).
Note that our  model is very sensitive to the parameter $\alpha$. Even for $(1-\alpha)\sim 10^{-3}-10^{-4}$ (i.e., almost Gaussian initial
conditions), the curves diverge significantly
from the strict Gaussian cases. This means that even very small deviations
from Gaussianity may introduce a significant change in the
cluster abundance. Actually, the presence of the second field tends to
slow down the cluster abundance evolution at high redshifts. 
In the case of $\Omega_m=1.0$ this effect is dramatic for $z>0.3$, while in
the case of $\Omega_m=0.2$ the difference is less pronounced and it
is clearer for $z\gtrsim 0.6$. Indeed, by plotting the 68\% confidence limits around the curve $\Omega_m=0.2$,  we see that Gaussian and non-Gaussian models are not clearly distinguishable for $z\leq 1$ (see Figure 1b). This 
is associated to the small number of observational points
and, possibly, to the simplicity of our model. However, even considering these caveats, our results seem to indicate that  small deviations from the strict Gaussianity may play an important role in the cluster abundance evolution. 

\section{Summary and Discussion}

We presented the first results of a study 
concerning  small deviations from Gaussianity in
the primordial density field. Using very simple arguments,
we developed a model based on the combination of two 
random fields in order to take into account the non-Gaussianity effects.
This model is physically motivated in the context of hybrid 
models, as well as  in the framework of biased scenarios for structure formation.  The weighted combined field involves a parameter $\alpha$ which modulates the relative importance of its components. For $\alpha\approx 1$,
we preserve the amplitude and shape of $P(k)$ and $\sigma(R)$ almost the same
as in the Gaussian case. At the same time, our results suggest that
even very small values of $(1-\alpha)$ can introduce a significant change in the cluster abundance evolution. This effect seems to be stronger in high density universes (at $z\leq 1$) than in low density universes where the effect probably turns more important at higher redshifts.

The model has some drawbacks. Firstly, it  depends on
the choice and amplitude of the second component of the combined field. Our choice of the lognormal function had a mathematical criterion of simplicity.
A detailed investigation of the use of different distribution
functions as the second component will be the subject of 
future works. However, the reasonable agreement between
the model and the data gives some support
to our arbitrary choice. Other possible limitation of this
work comes  from the use of the analytical approximation
to the density of non-linear objects following Sasaki (1994).
A more accurate description of the cluster abundance evolution 
requires the utilization of numerical methods. But 
Blain \& Longair (1993), also working in the Press \& Schecter
framework, found results numerically similar to Sasaki's, so
it seems that using this analytical approximation does not
introduce any systematical error. Finally, we should keep in mind
that our results are preliminary and both theoretical and 
observational efforts  are necessary in order to confirm or disproof
the hypothesis that the primordial density field can be described
as a slightly non-Gaussian distribution.

\acknowledgments{ We thank the anonymous referee for useful
suggestions. A.L.B. Ribeiro and P.S. Letelier thank the support of FAPESP.  P.S. Letelier also thanks the support of CNPq.}

\eject

\vspace{0.2cm}
\begin{figure}[h]
\begin{center}
\leavevmode
\epsfxsize=11.9cm
\epsfysize=11.9cm
\epsfbox{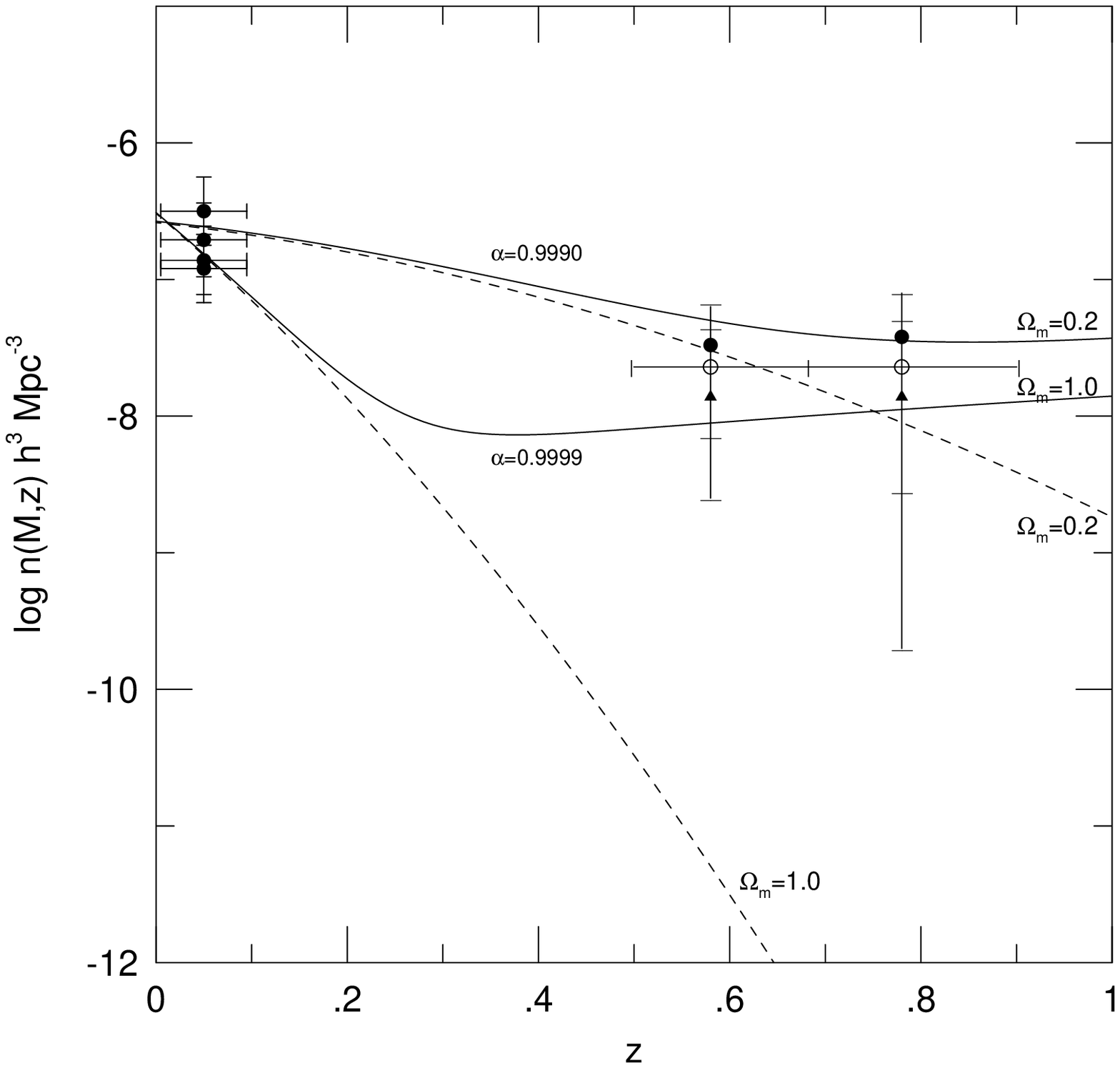}
\end{center}
\end{figure}
\vspace{0.1cm}
\noindent {\bf Fig. 1a -} Galaxy cluster abundance in the two-component
model for $\Omega_m=0.2$ with $\alpha=0.9990$ (solid) and $\alpha=1$
(dashed) and for $\Omega_m=1.0$ with $\alpha=0.9999$ (solid) and $\alpha=1$
(dashed). The curves, normalized at
$z=0$,  correspond to $n=-1.0$, $M_{\ast}=10^{14}~h^{-1}~M_{\sun}$
and $M>8\times 10^{14}~h^{-1}~M_{\sun}$. The observational
points were taken from Bahcall \& Fan (1998).

\eject
\vspace{0.2cm}
\begin{figure}[h]
\begin{center}
\leavevmode
\epsfxsize=11.9cm
\epsfysize=11.9cm
\epsfbox{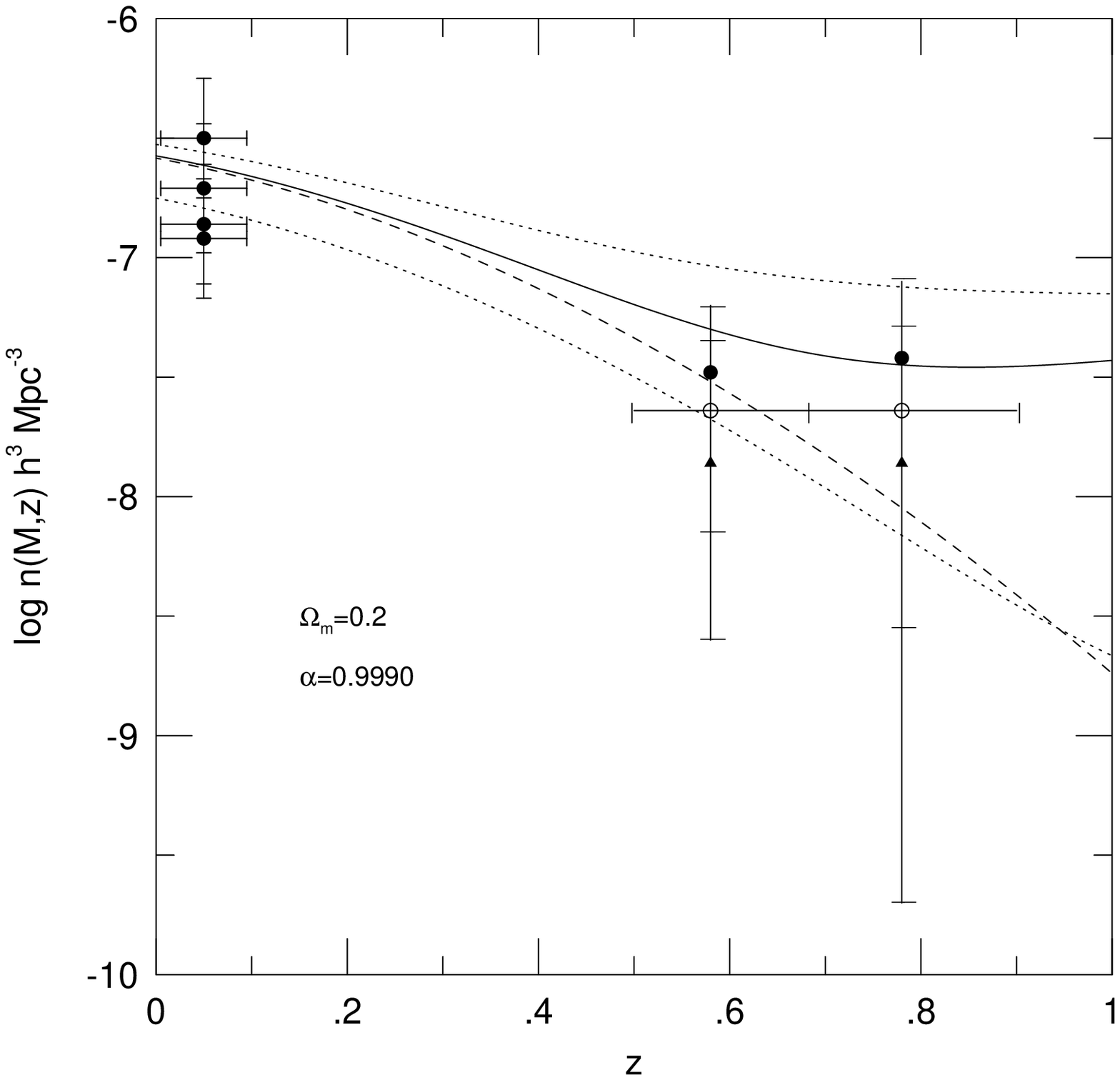}
\end{center}
\end{figure}
\vspace{0.1 cm}
\noindent {\bf Fig. 1b -}
Galaxy cluster abundance in the two-component
model for $\Omega_m=0.2$ with $\alpha=0.9990$ (solid) and $\alpha=1$
(dashed). The dotted lines are the 68\% confidence limits around the
non-Gaussian fit. The observational
points were taken from Bahcall \& Fan (1998).


\begin{references}
\reference{} Bahcall, N. 1999, astro-ph/9901076
\reference{} Bahcall, N. and Fan, X. 1998, ApJ, 504,1
\reference{} Bardeen, J.M., Bond, J.R., Kaiser, N. and Szalay, A.S. 1986, ApJ, 304, 15
\reference{} Battye, R.A. and Weller, J. 1998, In The
19th Texas Symposium on Relativistic Astrophysics and Cosmology, held in Paris, France, Dec. 14-18, 1998. Eds.: J. Paul, T. Montmerle, and E. Aubourg (CEA Saclay)
\reference{} Bi, H. and Davidsen, A.F., 1997, ApJ, 479, 523
\reference{} Blain, A.W. and Longair, M.S. 1993, MNRAS, 264, 509
\reference{} Coles, S. and Jones, B. 1991, MNRAS, 248, 1
\reference{} Ferreira, P., Magueijo, J. and G\'orski, K.M.G. 1998, ApJ, 503, L1
\reference{} Guth, A.H. and Pi, S.-Y. 1982, Phys. Rev. Lett., 49, 1110
\reference{} Henry, J.P. and Arnaud, K.A. 1991, ApJ, 372, 410
\reference{} Jensen, L.G. and Szalay, A.S. 1986, ApJ, 305, L5
\reference{} Kibble, T.W.B. 1976, J. Phys. A, 9, 1387
\reference{} Koyama, K., Soda, J. and Taruya, A. 1999, MNRAS, in press
\reference{} Magueijo, J. 1999, astro-ph/9911334
\reference{} Moscardini, L., Matarrese, S., Lucchin, F. and
Messina, A. 1991, MNRAS, 248, 424
\reference{} Ostriker, J.P. and Cowie, L.L. 1981, ApJ, 243, L127
\reference{} Pando, J., Valls-Gabaud, D. and Fang, L.Z. 1998, Phys. Rev. Lett. 81, 4568
\reference{} Pen, U.L. 1998, ApJ, 498, 60
\reference{} Plionis, M. and Valdarini, R. 1995, MNRAS, 272, 869
\reference{} Politzer, D. and Wise, M. 1984, ApJ, 285, L1
\reference{} Press, W.H. and Schechter, P. 1974, ApJ, 187, 425
\reference{} Salopek, D.S., Bond, J.R. and Bardeen, J.M. 1989, Phys. Rev. D,
40, 6
\reference{} Sasaki, S. 1994, PASJ, 46, 427
\reference{} Scherrer, R.J. and Bertschinger, E. 1991, ApJ, 381, 349
\reference{} Titterington, D.M., Smith, A.F.M. and Makov, U.E. 1985,
in Statistical Analysis of Finite Mixture Distributions,
Wiley Series in Probability and Mathematical Statistics, ed. John Wiley
\& Sons
\reference{} Weinberg, D.H. and Cole, S. 1992, MNRAS, 259, 652


\end{references}
\end{document}